\documentclass[12pt,showpacs,aps]{revtex4}

\usepackage{epsfig}

\def\cal#1{{\cal #1}}

\def\m@th{\mathsurround=0pt}
\def\n@space{\nulldelimiterspace=0pt \m@th}
\def\biggg#1{{\mbox{$\left#1\vbox to 20.5pt{}\right.\n@space$}}}

\def\beginenum{\begin{enumerate}}
\def\endenum{\end{enumerate}}
\def\bitem{\begin{itemize}}
\def\eitem{\end{itemize}}
\def\bray{\begin{array}}
\def\eray{\end{array}}
\def\begindoc{\begin{document}}
\def\enddoc{\end{document}}
\def\bq{\begin{equation}}
\def\eq{\end{equation}}
\def\bqy{\begin{eqnarray}}
\def\eqy{\end{eqnarray}}
\def\bqyn{\begin{eqnarray*}}
\def\eqyn{\end{eqnarray*}}
\def\bc{\begin{center}}
\def\ec{\end{center}}
\def\bfll{\begin{flushleft}}
\def\efll{\end{flushleft}}
\def\bflr{\begin{flushright}}
\def\eflr{\end{flushright}}
\newcommand{\Avec}{\mbox{\boldmath $A$}}
\newcommand{\Bvec}{\mbox{\boldmath $B$}}
\newcommand{\Evec}{\mbox{\boldmath $E$}}
\newcommand{\Fvec}{\mbox{\boldmath $F$}}
\newcommand{\Gvec}{\mbox{\boldmath $G$}}
\newcommand{\Rvec}{\mbox{\boldmath $R$}}
\newcommand{\Uvec}{\mbox{\boldmath $U$}}
\newcommand{\Vvec}{\mbox{\boldmath $V$}}
\newcommand{\evec}{\mbox{\boldmath $e$}}
\newcommand{\jvec}{\mbox{\boldmath $j$}}
\newcommand{\kvec}{\mbox{\boldmath $k$}}
\newcommand{\nvec}{\mbox{\boldmath $n$}}
\newcommand{\uvec}{\mbox{\boldmath $u$}}
\newcommand{\vvec}{\mbox{\boldmath $v$}}
\newcommand{\wvec}{\mbox{\boldmath $w$}}
\newcommand{\xvec}{\mbox{\boldmath $x$}}
\newcommand{\omegavec}{\mbox{\boldmath $\omega$}}
\newcommand{\Omegavec}{\mbox{\boldmath $\Omega$}}

\begin{document}

\title{Stable Optical Vortex Solitons in Pair Plasmas}
\author{V. I. Berezhiani}
\affiliation{Andronikashvili Institute of
Physics, Tbilisi 0177, Georgia}
\author{S. M. Mahajan}
\email{mahajan@mail.utexas.edu} \affiliation{Institute for Fusion
Studies, The University of Texas at Austin, Austin,Tx 78712}
\author{N. L. Shatashvili}
\email{shatash@ictp.it} \affiliation{Faculty of Exact and Natural
Sciences, Javakhishvili Tbilisi State
University, Tbilisi 0128, Georgia\\
Andronikashvili Institute of Physics, Tbilisi 0177, Georgia}

\begin{abstract}
{It is shown that the pair plasmas with small temperature
asymmetry can support existence of localized as well as
de-localized optical vortex solitons. Coexistence of such solitons
is possible due to peculiar form of saturating nonlinearity which
has a focusing-defocusing nature -- for weak amplitudes being
focusing becoming defocusing for higher amplitudes. It is shown
that delocalized vortex soliton is stable in entire region of its
existence while single- and multi-charged localized vortex
solitons are unstable for low amplitudes and become stable for
relativistic amplitudes. }
\end{abstract}

 \maketitle



\clearpage






\section{Model}


The richness of an electromagnetically active medium is often
measured by the variety of structures that it can support. Such
structures, in turn, are created because of the nonlinear response
of the medium, for instance, to the impact of a large amplitude
electromagnetic wave. Naturally the properties of the structure
(its  shape, its content, its stability, its angular momentum
etc.) are dictated by the type of nonlinearity that can arise in
the medium. The discovery or identification of a new nonlinearity
type, then, opens up a new era of investigation, even, discovery.

In this paper we work out some of the consequences of a new
focusing-defocusing nonlinearity \cite{bib:MSB} belonging to the
general class of saturating nonlinearities (whose magnitude tends
to a constant as the wave amplitude becomes large). Saturating
nonlinearities seem to appear, inter alia, in theories of large
amplitude wave propagation  in pair plasmas (plasmas whose main
constituents have equal mass and opposite charge
\cite{bib:sturrock,bib:blandford,bib:exp-ep}) in which the pair
symmetry is broken by some physical mechanism. For instance, a
small amount of Baryonic matter (protons) may break the symmetry
of an electron-positron (e-p) plasma in the MEV era of the early
universe \cite{bib:Wineberg,bib:shukla-ep,bib:BM,bib:theory-ep}.
In recently created pair ion (PI) plasmas in the laboratory, a
variety of symmetry breaking mechanisms like the small
contamination by a much heavier immobile ion, or a small mass
difference between the two constituent species, could exist
\cite{bib:OH1,bib:OH3,bib:OH4,bib:contamin,bib:SJK}. Asymmetries
originating in small temperature differences in the constituent
species may be always available for structure formation: in the
laboratory such a temperature difference could be readily
engineered (in a controlled way) and there are reasons to believe
that species temperature difference could exist in cosmic and
astrophysical setting where one encounters e-p plasmas. It is in
this latter setting that a new type of nonlinearity
\begin{equation}
F(\,|A|{\,^{2}})=\frac{\epsilon^{2}}{2}\,\frac{\kappa \,|A|{\,^{2}}}{\left( {1+\kappa
\,|A|^{\,2}}\right)^{2}}   \label{v01}
\end{equation}
appeared while deriving the wave equation (in parabolic
approximation) \cite{bib:MSB}
\begin{equation}
2i\omega _{0}\frac{\partial A}{\partial t}+\frac{(2-\epsilon )}{\omega
_{0}^{2}}\ \frac{\partial ^{\,2}A}{\partial \xi ^{2}}+\nabla _{\perp
}^{2}\,A+
F(\,|A|{\,^{2}})\cdot \,A=0 \ ,  \label{v1}
\end{equation}
describing the nonlinear evolution of the vector potential of an
electromagnetic pulse propagating in an arbitrary pair plasma with
temperature asymmetry. Following assumptions and notations are
necessary in order to put equations (\ref{v01}) and (\ref{v1}) in
perspective: \ $A$ \ is the slowly varying amplitude of the
circularly polarized EM pulse \ $\sim A\,(\hat{\bf x}+\hat{\bf
y})\,\rm{exp}(ik_0z-\omega_0t)$ \ with mean frequency $\omega_0$
and mean wave number $k_0$; \ $\nabla _{\perp }^{2}=\partial
^{2}/\partial x^{2}+\partial ^{2}/\partial y^{2}$ \ is the
diffraction operator and $\xi =z-v_{g}t$ is the "comoving" (with
group velocity $v_{g}$) coordinate.

Equation (\ref{v1}) is written in terms of the dimensionless
quantities \ $A=|e|A/(mG(T_{0}^{-})c^{2})$, \ $r=(\omega
_{e}/c)r$, \ $t=\omega_{e}t$; where
$\omega _{e}=(4\pi e^{2}n_{0}/m)^{1/2}$ is the electron Langmuir
frequency and $m$ is the electron mass. The charges $q^{\pm}$ and
masses $m^{\pm}$ of positive and negative ions are assumed to be
same (in this paper we mainly concentrate on the specific case of
pair plasma consisting of electrons and positrons, i.e.
$q^+=e^{+}=q^-=-e^{-}=|e|$ and $m^{+}=m^{-}=m$). The equilibrium
state of the system is characterized by an overall charge
neutrality $n_{0}^{+}=n_{0}^{-}=n_{0}$ where $n_{0}^{+}$ and
$n_{0}^{-}$ are the unperturbed number densities of the positive
and negative ions respectively. The background temperatures of
plasma species are $T_{0}^{\pm }$ ($T_{0}^{+}\neq T_{0}^{-}$) and
$m\,G(z^{\pm })=m\,K_{3}(z^{\pm })/K_{2}(z^{\pm })$ is the
"effective mass", [$z^{\pm }=mc^{2}/T^{\pm }$], where $K_{\nu }$
are the modified Bessel functions. For the nonrelativistic
temperatures ($T^{\pm }\ll mc^{2}$) $G^{\pm}=1+5T^{\pm }/2mc^{2}$
and for the ultra-relativistic temperatures ($T^{\pm}\gg m_{\alpha
}c^{2}$) $G^{\pm}=4T^{\pm }/mc^{2}\gg 1$. The smallness parameter
$\epsilon =[G(T_{0}^{+})-G(T_{0}^{-})]/G(T_{0}^{+})$ measures the
temperature asymmetry of plasma species. For the nonrelativistic
temperatures $\epsilon =5(T_{0}^{+}-T_{0}^{-})/2mc^{2}$ while in
ultrarelativistic case $\epsilon
=(T_{0}^{+}-T_{0}^{-})/T_{0}^{+}$. The numerical factor
$\kappa=1/2 $ for non-relativistic temperatures ($=2/3$ for
ultrarelativistic temperatures). In deriving Eq.(\ref{v1}) with
(\ref{v01}), we have assumed that the plasma is transparent
($\omega _{0}\gg 1$, $v_{g}\simeq 1$), and that the longitudinal
extent of the pulse is much shorter than its transverse
dimensions. However, despite of \ $\partial A/\partial \xi \gg
\nabla _{\perp }A$ , the second and the third terms in
Eq.(\ref{v1}) can be comparable due to the transparency of the
plasma ($\omega _{0}\gg 1$).

With self-evident renormalization the equation (\ref{v1}) can be
written as:
\begin{equation}
i\frac{\partial A}{\partial t}+\frac{\partial ^{\,2}A}{\partial \xi ^{2}}%
+\nabla _{\perp }^{2}\,A+f(|A|^{2})\cdot A=0 ,  \label{v2}
\end{equation}
where the nonlinearity function is now following \cite{bib:MSB}:
\begin{equation}
f(|A|^{2})=\frac{|A|^{2}}{(1+|A|^{2})^{2}}\ .  \label{v3}
\end{equation}
which has an unusual feature -- in the ultrarelativistic limit \
($|A|^{2}\gg 1$) \ it tends to be $0$.

Note that the nonlinear refraction index for the considered system
can be written as $\delta n_{nl}=f(I)$, where \ $I=|A|^{2}$ \ is
the intensity of the EM field. The medium is a self-focusing \
($d(\delta n_{nl})/dI>0$) \ provided $I<1$ while for higher
intensities ($I>1$), it becomes defocusing \ ($d(\delta
n_{nl})/dI<0$). For the localized intense EM pulse with the peak
intensity $I_{m}>1$ the medium becomes defocusing at the peak
while remaining focusing at the wings of the EM pulse intensity
profile.

In \cite{bib:MSB} we have demonstrated that Eq.(\ref{v2}) supports
existence of the stable solitonic structures for any spatial
dimensions $(D=1,2,3)$. Such "light-bullets" exist provided that
the amplitude of the solitons is lower than certain critical
values (for instance, in 1--dimensional ($1D$) media
$A_{m}<A_{mcr}\simeq 1.4)$. It is important to emphasize that at
$A_{m}\to A_{mcr}\ $the profile of the central part of the soliton
flattens and widens at the top. The existence of flat-top soliton
can be explained by the peculiarities of our focusing-defocusing
nonlinearity implying that the pulse top part with $A>1$ entered
the defocusing region having the tendency of diffraction while the
wings of the soliton are in focusing region preventing the total
spread of the pulse.

\section{Formation of Vortices}

In this section we examine the possibility of the formation of
two-dimensional stable soliton-structures carrying a screw type of
dislocation, i.e., optical vortices. The generation, propagation,
and interaction of optical vortices in nonlinear media has been a
subject of extensive studies (see for review \cite{bib:kivshar}).
In a self-defocusing medium an optical vortex soliton (OVS) is
($2+1$)--dimensional (two transverse dimension and a time)
stationary beam structure with phase singularity. An OVS is a dark
spot, i.e., a zero intensity center surrounded by a bright
infinite background. Self-focusing media also support localized
optical vortex soliton solutions (LOVS) with phase dislocation
surrounded by the bright ring. In self-focusing medium LOVS are
unstable against symmetry breaking perturbations that lead to the
breakup of rings into filaments \cite{bib:skryabin}.

Our nonlinearity (\ref{v3}) has focusing-defocusing features,
hence, one could expect that formation of both OVS and LOVS
solutions is possible in the medium. Such statement can be
augmented by the results of \cite{bib:BSA} where
focusing-defocusing model of the media was postulated to be
cubic-quintic medium with sign-changing nonlinearity
($f(|A|^2)=|A|^2-|A|^4$). In contrast to the cubic-quintic models
the saturation nonlinearity derived by us in \cite{bib:MSB} is not
sign-changing -- it is the focusing-defocusing one. To verify this
expectation we assume that the pulse is sufficiently long and
effects related to the group velocity dispersion ($\sim \omega
_{0}^{-2}\ \partial ^{\,2}A/\partial \xi ^{2}$) can be ignored.

\bigskip

Introducing polar coordinates ($r,$ $\theta $) in ($x,y$) plane,
we look for solutions of Eq.(\ref{v2}) in the form of
\begin{equation}
A=A(r)\exp (i\lambda t+im\theta ) ,    \label{v4}
\end{equation}
where integer \ $m$ \ defines the topological charge of vortices
and \ $\lambda $ \ is the nonlinear frequency shift. The ansatz
(\ref{v4}) converts Eq. (\ref {v2}) to the ordinary differential
equation
\begin{equation}
\frac{d^{2}A}{dr^{2}}+\frac{1}{r}\frac{dA}{dr}-\frac{m^{2}}{r^{2}}A-\lambda
A+\frac{A^{3}}{(1+A^{2})^{2}}=0  .   \label{v5}
\end{equation}

We have used numerical methods to find the localized solutions of
(\ref {v5}). It is possible to map the equation in the ($A,A_{r}$)
plane (phase plane) and show that it admits both OVS and LOVS
solutions. LOVS can exist in the form of infinite number of
discrete bound states with $A_{n}(r)$ ($n=1,2,...$) where $%
n$ denotes the finite $r$ zeros of the eigenfunction.

\bigskip

In what follows we consider only the lowest order (lowest radial
eigenmode) solution of Eq.(\ref{v5}) ($n=1$). For nonzero \ $m$ \
(the case we are interested in), the ground state LOVS is
positive, has a node at the origin $r=0$, reaching a maximum, and
then monotonically decreases with increasing $r$. Such localized
solution exists if $\lambda >0$ with the following asymptotic
behavior: \ $A_{r\rightarrow 0}\rightarrow r^{|m|}A_{0}$ \ and \
$A_{r\rightarrow \infty }\rightarrow \exp (-r\sqrt{\lambda
})/\sqrt{r}$ , where \ $A_{0}$ \ is a constant which measures the
slope of $A$ at origin. OVS solutions have the same asymptote for
\ $r\rightarrow 0$, while for \ $r\rightarrow \infty $ \ the
amplitude has a nonzero value \
$A(r)=A_{\infty}+m^{2}/(r^{2}f^{\prime }(A_{\infty }))$ . Here, \
$\lambda =f(A_{\infty })$ \ and OVS exists when
$f^{\prime}(A_{\infty })<0$ , i.e. when background intensity of
the soliton (far beyond the vortex core) is relativistic \
$A_{\infty }>1$. In dimensional units this condition corresponds
to the negative slope of the nonlinear refractive index \
($d\delta n_{nl}/dI<0$), \ i.e., in the asymptotic region of the
solution the medium is defocusing. It is easy to demonstrate
\cite{bib:MSB} that the constant background field with \
$A_{\infty }>1$ \ is modulationally stable.

A shooting code was used to numerically solve Eq.(\ref{v5}). To
better understand the results of simulations we use the analogy
with a nonconservative motion of a particle. For this purpose one
can rewrite the Eq.(\ref{v5}) as
\begin{equation}
\frac{d}{dr}\left[ \left( \frac{dA}{dr}\right) ^{2}+V(A)\right]
=\frac{m^{2}}{r^{2}}\frac{dA^{2}}{dr}-\frac{2}{r}\left(
\frac{dA}{dr}\right) ^{2} ,   \label{v6}
\end{equation}
where the "effective potential" is \ $V(A)=-\lambda A^{2}+\ln
(1+A^{2})-A^{2}/(1+A^{2})$ . The profile of the potential \ $V(A)$
\ for different values of \ $\lambda $ \ is presented in Fig.1.
The potential has the maxima at the points \ $A=0$ \ and \
$A_{\max }=\sqrt{\left[ 1-2\lambda +\sqrt{1-4\lambda }\right]
/2\lambda }$ . The bounded solution is possible only in the case \
$0<\lambda <0.25$ \ while \ $A_{\max }>1$.

The OVS solutions correspond to a particle beginning its motion at
origin ($A=0$) with certain initial $A_{0}$ (which can be termed
as a velocity (if $m=1$) or acceleration (if $m=2$ and so on) and
dissipating its initial energy approaching asymptotically
potential maximum at $A_{\max }$. Thus, the background intensity
of OVS is $A_{\infty }=A_{\max }$, it is always larger than unity
and can become arbitrarily large for $\lambda \rightarrow 0$. We
also found out that OVS solutions exist even for $0.25>\lambda
>\lambda _{cr}\simeq 0.2162$, i.e., when $V(A_{\max})<0$
(see curve "a" in Fig.1). In other words, the effective particle
cannot overpass but only approach asymptotically the lower
potential maximum.

The numerical solutions of the nonlinear equation (\ref{v5}) for
$m=1,2$ and $3$ are shown in Fig.2. As expected, the soliton-like
solutions evidently go to zero as \ $r^{m}$ \ for small $r$, and
reach an $m$-independent asymptotic value predicted above. In
Fig.3 curve "a" displays the dependence of the field derivative at
the origin, \ $A_{0}$ , \ as a function of the nonlinear frequency
shift \ $\lambda$ \ for \ $m=1$ \ case. One can see that \ $A_{0}$
\ is a growing function for small \ $\lambda $-s . For small
$\lambda $-s the position of the potential maximum "moves" to
larger values of $A$ and, consequently, "particle" needs to have
larger initial "velocity" ($A_{0}$) to reach the maximum.

In contrast to OVS, the LOVS solutions correspond to the particle
returning back asymptotically to the initial position at $A=0$. It
seems obvious that due to the "frictional" motion particle can not
make its way back if $\lambda >\lambda _{cr}$. Thus, LOVS may
exist in the range $0<\lambda <\lambda _{cr}$ while its amplitude
(in contrast to OVS) is a growing function of $\lambda $. Such
dependence is obtained numerically and is presented in Fig.4. for
single-charged vortices ($m=1$). One can see that the amplitude of
the LOVS ($A_{m}$) is bounded from above by certain critical value
for $A_{cr}(\simeq 1.5)$. Thus, in contrast to OVS the localized
vortex can be just moderately relativistic. Notice that for \
$0.16 \leq \lambda \leq \lambda _{cr}$ \ the amplitude of the LOVS
($A_{m}$) varies in the range \ $1\leq A_{m}\leq A_{cr}$ . For the
top part of such solution (with $A(r)>1$) the medium is defocusing
while remains focusing for lower intensity wings of the structure.
Consequences of this fact can be seen in Fig.5 where profile of
LOVS is given for variety of $\lambda $. With increase of $\lambda
$ the central part of the LOVS flattens and widens converging to
the OVS. In principle, it is possible to create flat-top LOVS with
a large transverse width. Convergence of LOVS to OVS can be better
seen in Fig.3 where the curve "b" corresponding to LOVS almost
coincides with curve "a" near the point $\lambda \approx \lambda
_{cr}$. Similar behavior of the solutions can be obtained for the
vortices with higher charge ($m=2,3,..$), however, the
corresponding figures we do not present here for brevity.

\section{Stability of Solutions}

Are these solitonlike solutions stable?

The intensity dependent switching from the focusing to defocusing
regime can have an interesting consequence for the stability
properties of the solutions. As it is well established
\cite{bib:kivshar} OVSs with $m=1$ are stable whereas vortices
with a larger value of \ $m$ \ may decay into the single-charged
ones in self-defocusing media. In our case the bulk of the OVS is
always in the defocusing regime and as we mentioned above the
background field is always stable. However, near to vortex core
the medium becomes focusing. Thus, stability of the OVS in our
case can not be granted.

We performed stability analysis solving numerically Eq.(\ref{v2});
\ while simulation (for various $\lambda $-s),
initial stationary OVS state was perturbed radially and
azimuthally by Gaussian noise. Typical picture of the evolution is
plotted is Fig.6. We see that perturbations are quickly radiated
away and the initial state relaxes to the ground state solution
implying that the OVS is stable in the whole examined range.

To verify stability of LOVS we first performed a linear stability
analysis. To do so we follow procedure developed by
\cite{bib:berge} and consider perturbation acting along a ring of
mean radius \ $r_{*}$ , \ where \ $A(r_{*})=A_{m}$ . Assuming
constant intensity and spatial uniformity for this ring, one can
rewrite the diffraction operator in (\ref{v2}) as \ $\nabla _{\bot
}^{2}=r_{*}^{-2}\partial ^{2}/\partial \theta ^{2}$ \ and
introducing azimuthal perturbation with a phase factor \ $\Psi
=\Omega t+M\theta $ \ (where $M$ is an integer) for the growth
rate of instability we get:
\begin{equation}
\mathop{\rm Im} (\Omega )=\frac{M}{r_{*}} \mathop{\rm Re}
\left[ \frac{2(1-A_{m}^{2})}{(1+A_{m}^{2})^{3}}-\frac{M^{2}}{r_{*}^{2}}%
\right] .  \label{v7}
\end{equation}
One can see from (\ref{v7}) that large amplitude LOVS with \
$A_{m}>1$ \ is always stable. For the lower amplitude case LOVS
should decay into $M_{\max }$ multiple filaments, where \
$M_{\max}$ \ is an integer close to the number for which maximal
growth rate is maximal.

In Fig.7 we plot \ $\mathop{\rm Im}(\Omega )$ \ versus \ $M$ \ for
\ $\lambda =0.1$ \ and for \ $m=1,2,3$ . Corresponding \ $A_{m}$ \
are respectively $0.66$; \ $0.65$; \ $0.63$ \ and $r_{*}$=$6.3$; \
$11.6$; \ $16.9$. One should expect that instability will result
in breaking of the pulse into the filaments (fragments) with
number of filaments being respectively 2, 4, and 5 (or 6) for \
$m=1,2,3$. These filaments have to conserve the total angular
momentum. Since the fusion of filaments is not possible due to the
topological reasons, they can eventually spiral about each other
or fly off tangentially to the initial ring generating bright
solitonic structures found for index saturation nonlinearity
\cite{bib:skryabin}.

\bigskip

Our numerical simulations for $A_{m}<1$ give evidence of a quickly
developing instability in agreement with predictions of linear
stability analysis. Indeed, in Fig.8 we present the results of
simulations for the LOVS with $m=1$ and $2$. The LOVS\ with $m=1$
breaks into two filaments while for \ $m=2$ \ the breaking into
$4$ filaments takes place. The filaments are running away
tangentially without spiraling. All filaments like spatial
solitons remain stable. Most interesting is the situation when
amplitude of LOVS is larger than unity.

In Fig.9 we present the evolution of LOVS both for \ $m=1$ \ and \
$m=2$ ; corresponding amplitudes for the soliton solutions are \
($\lambda =0.2$)\ : \ $A_m=1.39$ \ and \ $A_m=1.37$  ,
respectively. The initial input LOVS solution was modulated by a
Gaussian noise (the level of noise was $5\%$). One can see that
breaking of the LOVS does not take place. In order to be sure that
some very slow instability is not developing the simulations were
carried out until $t=4000$, i.e., for $130$ soliton period
$T_{sol}=2\pi /\lambda \approx 30$. Thus, single and multi-charged
LOVS become stable for large amplitudes.

\bigskip

At the end we would like to emphasize that if for single-charged
LOVS we are confident about its stability for multi-charge LOVS
(as well as for OVS) one should be careful. Indeed, from general
topological reasons the multi-charged vortices are supposed to be
unstable and they should break into single charge vortices.
However, we could not observe in our simulations such breaking. It
is possible that this instability is very slow
(sub-exponential/algebraic) and, as a result we obtained that
multi-charged vortices are very long-lived objects and practically
stable.

\bigskip

In our consideration the effects related to the group velocity
dispersion and corresponding reshaping of the radiation have been
ignored. However, one can generalize our results by keeping the
term $\sim \partial ^{\,2}A/\partial \xi ^{2}$ in Eq.(\ref{v2}).
In transparent plasma case this term can affect the long time
dynamics of self-guiding vortex solitons. In particular due to
weak modulation instability \cite{bib:akhmediev} the self-trapped
beam eventually will break into a train of spatiotemporal
solitons, i.e. the "light bullets". However, due to the
topological reasons the vortex lines should survive the structural
changes. We expect that instability will result in generation of
fully localized bullets of vortex solitons (the spinning-bullets).
Dynamics of formation and stability of such structures is beyond
of the scope of current paper.

\section{Conclusions}

The asymmetries originating in small temperature differences in
the constituent species of electromagnetically active medium may
be always available for structure formation both in laboratory and
cosmic/astrophysical settings. In present paper we have shown that
this asymmetry defines the specific properties of the structure
due to the new type of the nonlinearity  that can arise in the
medium (derived in our earlier paper \cite{bib:MSB}). We found
that the pair plasmas of any dimension with "asymmetry" in initial
temperatures can support the stable large amplitude optical vortex
and localized vortex solitons. Localized structures for certain
parameters may have the flat-top shapes. The coexistence of LOVS
and OVS solutions and their stability in such medium is due to the
specific form of saturating nonlinearity switching from the
self-focusing to the self-defocusing regime and vice versa. This
fact is rather interesting specifically for laboratory conditions.

\bigskip

\centerline{***}

Authors express their special thanks to Doctors S.I. Mikeladze and
K.I. Sigua for the interest. The work of SMM was supported by
USDOE Contract No.~DE--FG 03-96ER-54366. The work of NLS and VIB
was supported by ISTC Project G-1366 and Georgian NSF grant
projects GNSF 69/07 (GNSF/ST06/4-057) and GNSF 195/07
(GNSF/ST07/4-191).

\clearpage

\vspace{2cm}


\begin{figure}
\begin{center}
\includegraphics[scale=0.33,angle=-90]{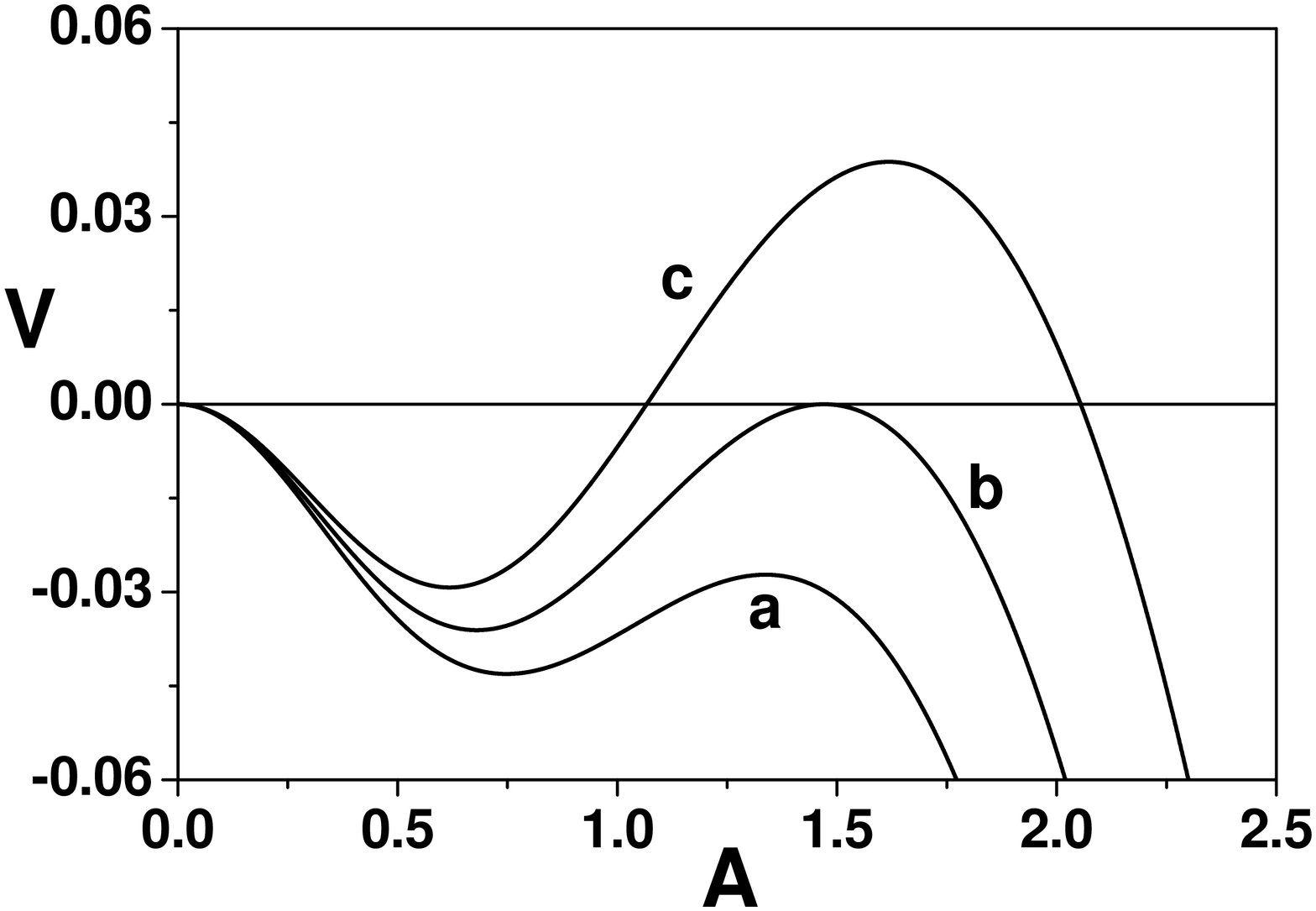}
\caption{ \ \ The "effective potential" versus the amplitude for
different values of the nonlinear frequency shift $\lambda $. The
curve "a" corresponds to $\lambda >\lambda_{cr}\simeq 0.2162$, the
curve "b" has $\lambda =\lambda _{cr}$, and for the curve "c"
$0<\lambda <\lambda _{cr}$ . }
\label{fig:potential}
\end{center}
\end{figure}

\begin{figure}
\begin{center}
\includegraphics[scale=0.33,angle=-90]{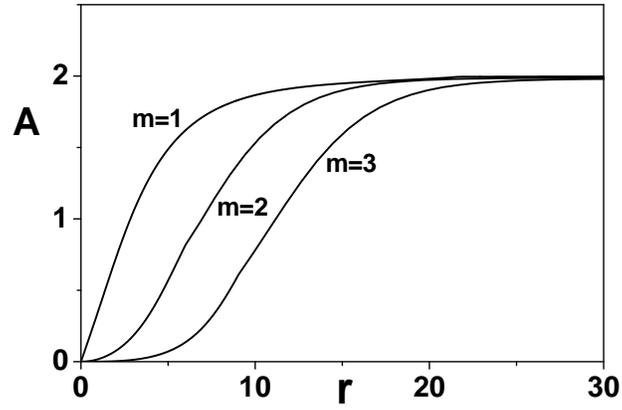}
\caption{ Profiles of OVS-s for $m=1, \ m=2, \ m=3$; \ nonlinear
frequency shift $\lambda=0.16$. }
\label{fig:profiles}
\end{center}
\end{figure}

\begin{figure}
\begin{center}
\includegraphics[scale=0.33,angle=-90]{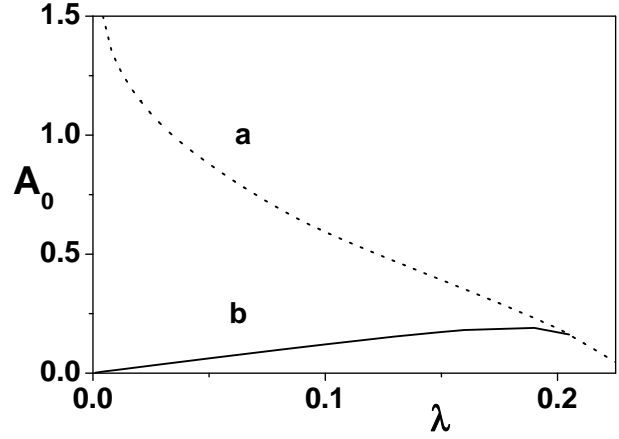}
\caption{$A_0$ versus $\lambda$ for $m=1$; curve "a" corresponds
to OVS while curve "b" -- to LOVS.}
\label{fig:A-1}
\end{center}
\end{figure}

\begin{figure}
\begin{center}
\includegraphics[scale=0.33,angle=-90]{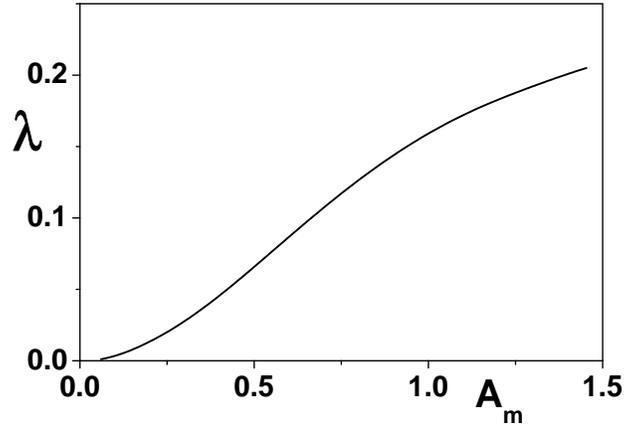}
\caption{The effective eigenvalue $\lambda$ versus soliton
amplitude $A_m$ for $m=1$.}
\label{fig:lambda}
\end{center}
\end{figure}

\begin{figure}
\begin{center}
\includegraphics[scale=0.33,angle=-90]{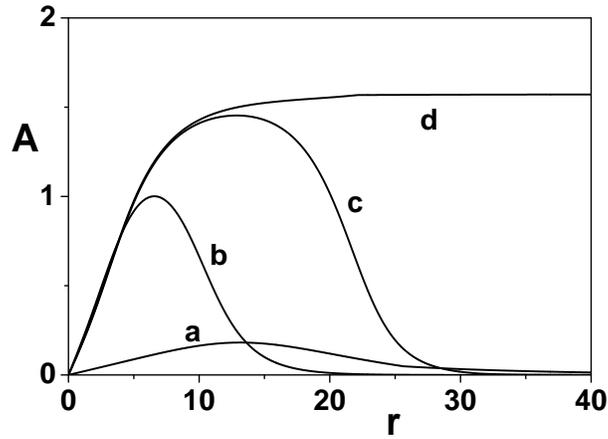}
\caption{ Profiles of soliton solutions. Curves "a", "b" and "c"
correspond to LOVS with $\lambda=0.005; \ 0.16; \ 0.205$,
respectively. Curve "d" corresponds to OVS for $\lambda=0.205$. }
\label{fig:solitons}
\end{center}
\end{figure}

\begin{figure}
\begin{center}
\includegraphics[scale=0.45,angle=-90]{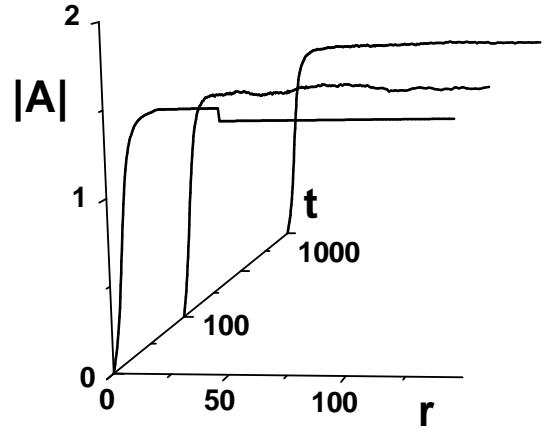}
\caption{ The dynamics of initially perturbed OVS-s: plots are
chosen for different time-moments
$t=0 ; \ 100 ; \ 1000$ .}
\label{fig:OVS-dynamics}
\end{center}
\end{figure}

\begin{figure}
\begin{center}
\includegraphics[scale=0.33,angle=-90]{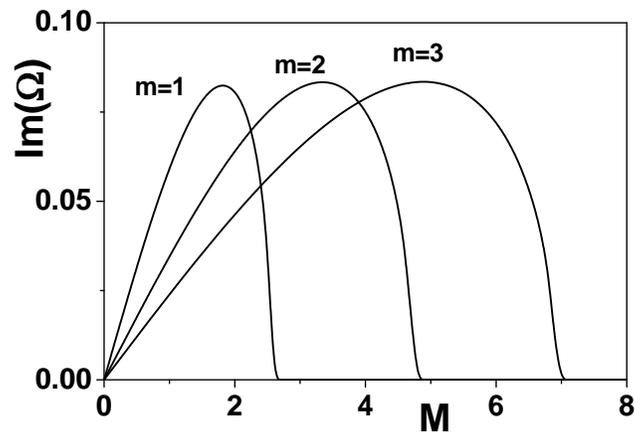}
\caption{Instability growth rate  $Im(\Omega)$ versus $M$ for
$\lambda=0.1$ for different topological charges $m$. }
\label{fig:instability}
\end{center}
\end{figure}

\begin{figure}
\begin{center}
\includegraphics[scale=0.5]{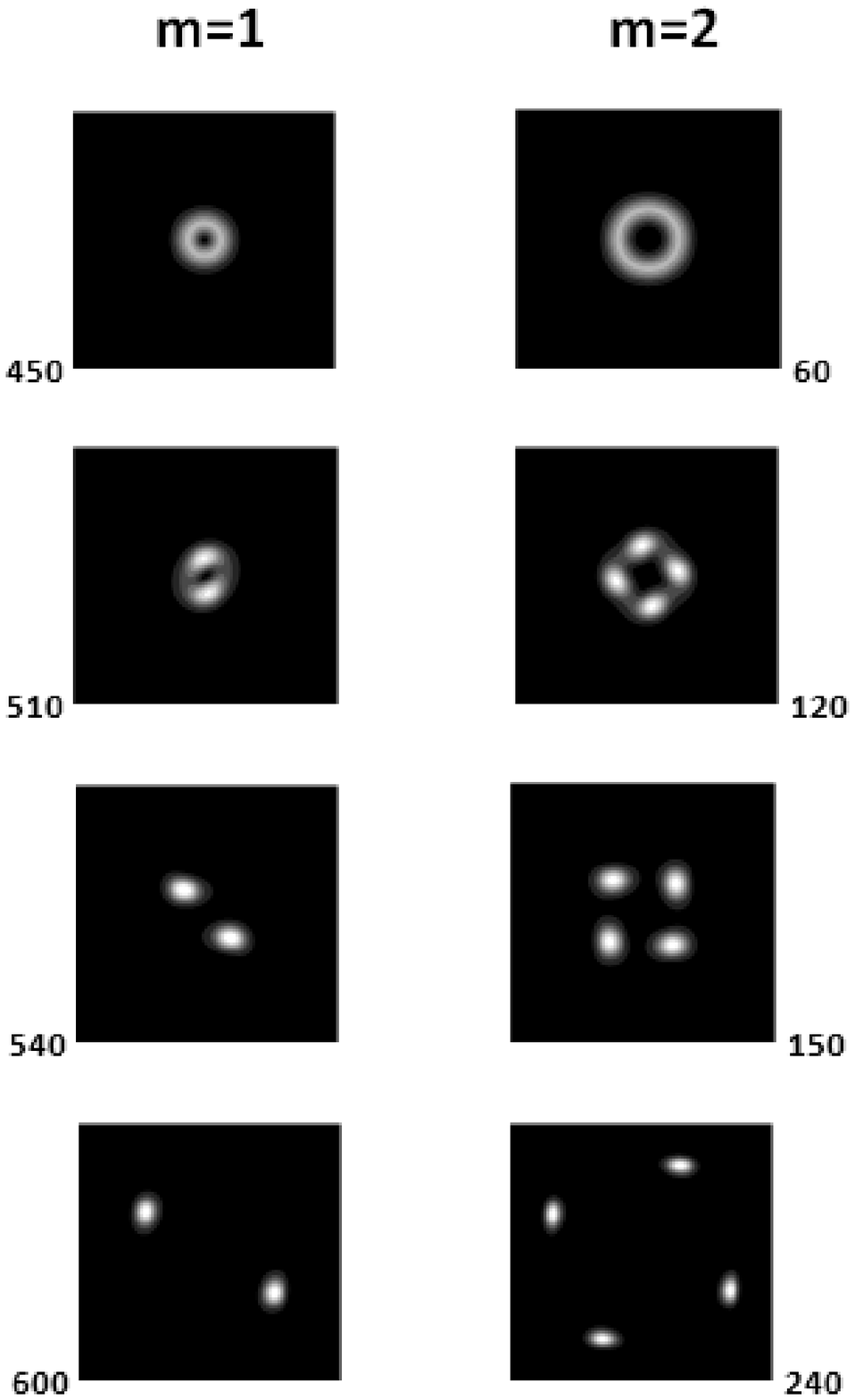}
\caption{Vortex dynamics (for different time-moments) when
$\lambda=0.1 $: the left panel -- for $m=1 , \ A_{max}=0.66 \ $,
the vortex splits into 2 filaments; the right panel -- for $m=2 ,
\ A_{max}=0.6580 $, the vortex splits into 4 filaments; the
filaments are running away tangentially.}
\label{fig:Vortex-dynamics}
\end{center}
\end{figure}

\begin{figure}
\begin{center}
\includegraphics[scale=0.5]{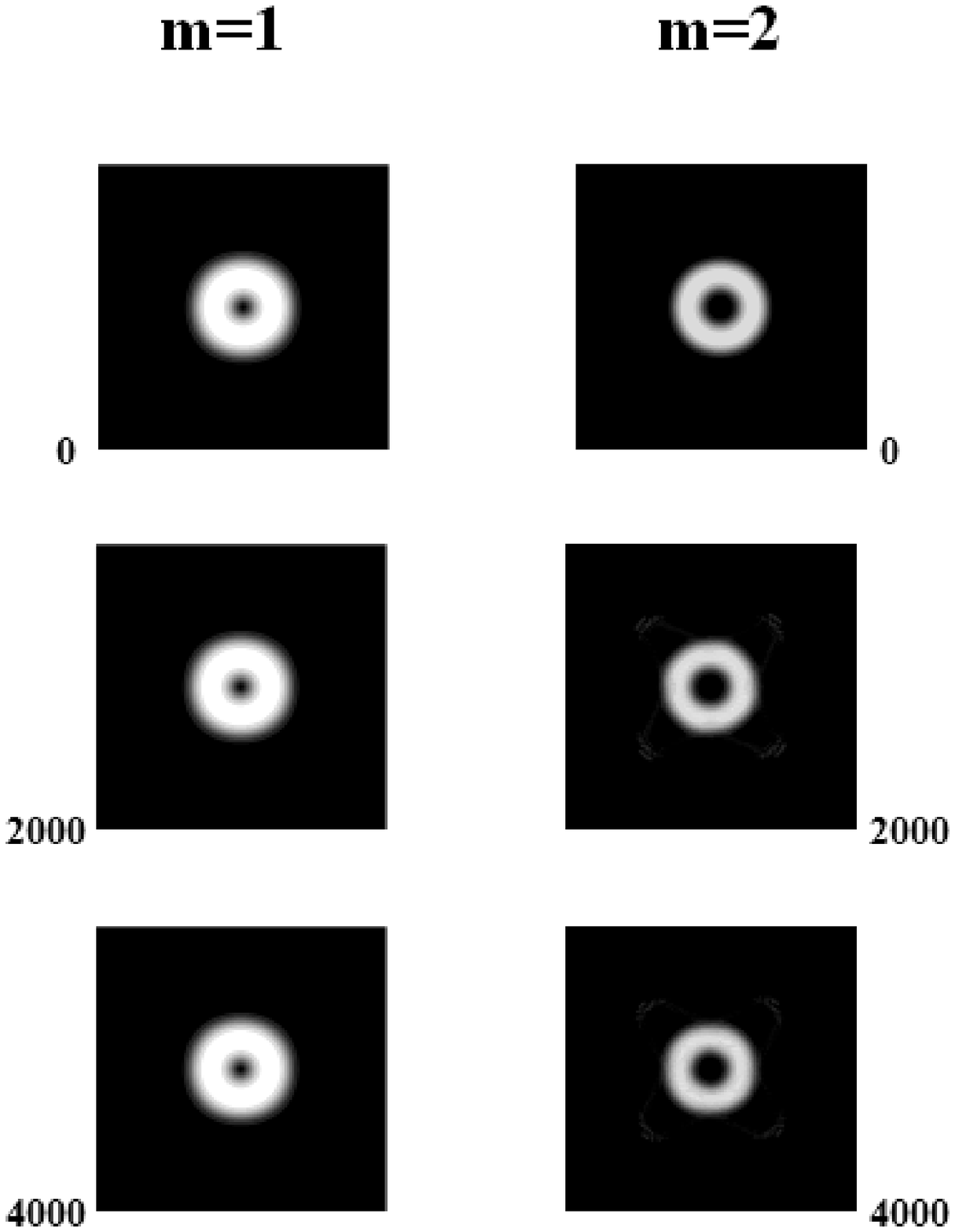}
\caption{Vortex dynamics (for different time-moments) when
$\lambda=0.2 $, the vortex is robust towards perturbations; the
left panel is for \ $m=1 , \ A_{max}=1.386 $ \ while the right
panel is for \ $m=2 , \ A_m=1.3729 $.} \label{fig:robust}
\end{center}
\end{figure}

\end{document}